\documentclass[aps, twocolumn,nofootinbib,prd]{revtex4}
\usepackage{amsmath}
\usepackage{graphicx}
\usepackage{graphics}
\usepackage{epsfig}

\def\simlt{\lower.5ex\hbox{$\; \buildrel < \over \sim \;$}}
\def\simgt{\lower.5ex\hbox{$\; \buildrel > \over \sim \;$}}
\def\simpropto{\lower.2ex\hbox{$\; \buildrel \propto \over \sim \;$}}

\newcommand{\newc}{\newcommand}
\newc{\gsim}{\lower.7ex\hbox{$\;\stackrel{\textstyle>}{\sim}\;$}}
\newc{\lsim}{\lower.7ex\hbox{$\;\stackrel{\textstyle<}{\sim}\;$}}
\newc{\gev}{\,{\rm GeV}}
\newc{\mev}{\,{\rm MeV}}
\newc{\ev}{\,{\rm eV}}
\newc{\kev}{\,{\rm keV}}
\newc{\tev}{\,{\rm TeV}}

\newc{\mz}{M_Z}
\newc{\mpl}{M_*}
\newc{\mw}{m_{\rm weak}}
\newc{\nr}[1]{N^c_R{}_{#1}}
\usepackage{amsmath}
\def\beq{\begin{equation}}
\def\eeq{\end{equation}}
\def\bea{\begin{eqnarray}}
\def\eea{\end{eqnarray}}
\def\bitem{\begin{itemize}}
\def\eitem{\end{itemize}}

\newc{\ie}{{\it i.e.}}          \newc{\etal}{{\it et al.}}
\newc{\eg}{{\it e.g.}}          \newc{\etc}{{\it etc.}}
\newc{\cf}{{\it c.f.}}
\def\bar#1{\overline{#1}}

\def\inv{^{\raise.15ex\hbox{${\scriptscriptstyle -}$}\kern-.05em 1}}
\def\lbar{{\lower.35ex\hbox{$\mathchar'26$}\mkern-10mu\lambda}} 

\def\to{\rightarrow}

\let\<=\langle
\let\>=\rangle

\let\+=\uparrow

\newcommand{\mdm}{m}
\newcommand{\mV}{m_{\mathrm{V}}}
\newcommand{\eps}{\varepsilon}
\newcommand{\GeV}{\mathrm{GeV}}
\newcommand{\TeV}{\mathrm{TeV}}
\newcommand{\kpc}{\mathrm{kpc}}

\begin{document}    
\title{{\bf\large{Can  the  WIMP annihilation boost factor be boosted by the Sommerfeld enhancement?}}} 
\author{Massimiliano Lattanzi}
\email{mxl@astro.ox.ac.uk}
\affiliation{Physics Department, University of Oxford, OX1 3RH Oxford, UK}
\affiliation{Istituto Nazionale di Fisica Nucleare, Via Enrico Fermi, 40 - 00044 Frascati, Rome, Italy}
\author{Joseph Silk}
\email{silk@astro.ox.ac.uk}
\affiliation{Physics Department, University of Oxford, OX1 3RH Oxford, UK}

\begin{abstract}
  We demonstrate that the Sommerfeld correction to cold dark matter (CDM) annihilations can be appreciable if even a small component of  the dark matter is extremely cold. Subhalo substructure provides such a possibility
given that the smallest clumps are relatively cold and contain even colder substructure  due to incomplete phase space mixing. Leptonic channels can be enhanced for plausible models and the solar neighbourhood boost required to account for PAMELA/ATIC data is plausibly obtained, especially in the case of  a few TeV mass neutralino for which the Sommerfeld-corrected boost is found to be $\sim10^4-10^5.$ Saturation of the Sommerfeld effect   is shown to occur below $\beta\sim 10^{-4},$  
thereby making this result largely independent on the presence of substructures below $\sim 10^5\rm M_\odot$. We find that  the associated diffuse gamma ray signal from annihilations  would  exceed EGRET constraints unless the channels 
annihilating to heavy quarks or to gauge bosons 
are suppressed. The lepton channel gamma rays are potentially detectable by the FERMI
satellite, not from the inner galaxy where substructures are tidally disrupted, but rather as a quasi-isotropic background from the outer halo, unless  the outer substructures are  much less concentrated than the inner substructures  and/or the CDM density profile  out to  the virial radius steepens significantly.
\end{abstract}


\maketitle 
\oddsidemargin=-.3in

\def\simlt{\lower.5ex\hbox{$\; \buildrel < \over \sim \;$}}
\def\simgt{\lower.5ex\hbox{$\; \buildrel > \over \sim \;$}}
\def\simpropto{\lower.2ex\hbox{$\; \buildrel \propto \over \sim \;$}}
\section{ Introduction}
The motivation for studying dark matter annihilation signatures 
(see e.g. \cite{Bertone:2004pz})
has received considerable recent attention following reports of a 100 GeV excess in the PAMELA data on the ratio of the fluxes of cosmic ray positrons to electrons \cite{Adriani:2008zr}.
 In the absence of any compelling astrophysical explanation, the signature is  reminiscent of the original prediction of a unique dark matter annihilation signal \cite{Silk:1984zy},
 although there are several  problems that demand attention before any definitive statements can be made.  By far the most serious of these is the required annihilation boost factor. The remaining difficulties with a dark matter interpretation, including most notably the gamma ray signals from the Galactic Centre and the inferred leptonic branching ratio, are,  as we argue below, plausibly circumvented or at least alleviated.
Recent data from the ATIC balloon experiment  provides evidence for a cut-off in the positron flux near 500 GeV that supports a Kaluza-Klein-like candidate for the annihilating particle \cite{chang:2008zzr} or a neutralino with incorporation of suitable radiative corrections \cite{berg}.

In a pioneering paper,  it was  noted \cite{Profumo:2005xd}
that the annihilation signal can be boosted by a combination of coannihilations and Sommerfeld corrrection. We remark  first that the inclusion of coannihilations to boost the annihilation cross-section modifies the relic density, and opens the 1-10 TeV  neutralino mass window to the observed (WMAP5-normalised) dark matter density. As found by  \cite{Lavalle et al.}, 
the outstanding problem now becomes  that of normalisation.  A boost factor of around  100 is required to explain the HEAT data in the context of a 100 GeV neutralino.  The flux is suppressed by between one and two powers of neutralino mass, and the problem becomes far more severe with the 1-10 TeV neutralino  required by the PAMELA/ATIC data \cite{Cirelli:2008pk}, a boost of $10^4$ or more being required. 
These latter authors included a Sommerfeld correction appropriate to our $\beta\equiv v/c=0.001$ dark halo and incorporated channel-dependent boost factors to fit the data, but  the required boosts still fell short of plausible values by  at least an order of magnitude. 

Here we propose a solution to the boost problem  via Sommerfeld correction in the presence of a model of substructure that incorporates a plausible phase space structure for cold dark matter (CDM). We reassess the difficulty with the  leptonic branching ratio and show that it is not insurmountable  for supersymmetric candidates. Finally, we evaluate  the possibility of independent confirmation via photon channels.

Substructure survival means that as much as 10\%  
of the dark matter is at  much lower $\beta$. This is likely in the solar neighbourhood and beyond, but not in the inner galaxy where clump destruction is prevalent due to tidal interactions.  Possible annihilation signatures from the innermost galaxy such as the WMAP haze  of synchrotron emission  and the EGRET flux of diffuse gamma rays  are likely to be much less affected by clumpy substructure than the positron flux in the solar neighbourhood.  We show in the following section that incorporation of the Sommerfeld correction means that clumps dominate the annihilation signal, to the extent that the initial clumpiness  of the dark halo survives. 

\section{The Sommerfeld enhancement}
\label{sec:som}

\begin{figure}
	\begin{center}
	\includegraphics[width=0.45\textwidth,  keepaspectratio]{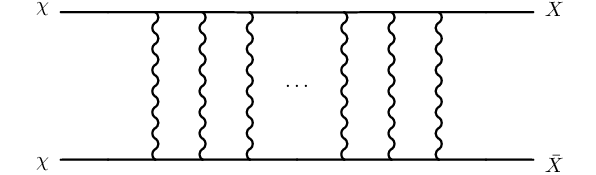}
\caption{Ladder diagram giving rise to the Sommerfeld enhancement for $\chi\chi\to X\bar X$ annihilation, via the exchange of gauge bosons. }
	\label{fig:somgen}
         \end{center}
\end{figure}

Dark matter annihilation cross sections in the low-velocity regime can be enhanced through the so-called ``Sommerfeld effect" \cite{somm31,hisano,Hisano05,Cirelli:2007xd, MarchRussell:2008yu,ArkaniHamed:2008qn,Pospelov08}.
This non-relativistic quantum effect arises because, when the particles interact through some kind of force, their wave function is distorted by the presence of a potential
if their kinetic energy is low enough. In the language of quantum field theory, this correspond to the contribution of ``ladder'' Feynman diagrams like the one shown in Fig. \ref{fig:somgen} in which the force carrier is exchanged many times before the annihilation finally occurs. This gives rise to (non-perturbative) corrections to the cross section for the process under consideration. The actual annihilation cross section times velocity will then be:
\begin{equation}
\sigma v = S \left(\sigma v\right)_0
\end{equation}
where $\left(\sigma v\right)_0$ is the tree level cross section times velocity, and in the following we will refer to the factor $S$ as the ``Sommerfeld boost'' or ``Sommerfeld enhancement'' \footnote{ In the case of repulsive forces, the Sommerfeld ``enhancement'' can actually be $S<1$, although we will not consider this possibility here.}. 

In this section we will study this process in a semi-quantitative way using a simple case, namely that of a particle interacting through a Yukawa potential.
We consider a dark matter particle of mass $\mdm$.
Let $\psi(r)$ be the reduced two-body wave function for the s-wave annihilation; in the non-relativistic limit, it will obey the radial Schr\"odinger equation:
\begin{equation}
\frac{1}{\mdm}\frac{d^2\psi(r)}{dr^2}-V(r)\psi(r)=-\mdm\beta^2\psi(r), 
\label{eq:sch}
\end{equation}
where $\beta$ is the velocity of the particle and  $V(r) = -\frac{\alpha}{r} e^{-\mV r}$ is an attractive Yukawa potential mediated by a boson of mass $\mV$.

The Sommerfeld enhancement $S$ can be calculated by solving the  Schr\"odinger equation with the boundary condition $d\psi/dr = i\mdm\beta\psi$ as $r\to\infty$. 
Eq. (\ref{eq:sch}) can be easily solved numerically. It is however useful to consider some particular limits in order to gain some qualitative insight into the dependence of the Sommerfeld enhancement on particle mass and velocity.  First of all, we note that for $\mV\to 0$, the potential becomes Coulomb-like. In this case the Schr\"odinger equation can be solved analytically; the resulting Sommerfeld enhancement is:
\begin{equation}
S=\frac{\pi\alpha}{\beta}({1-e^{-\pi\alpha/\beta}})^{-1}.
\end{equation} 
For very small velocities ($\beta\to 0$), the boost $S\simeq \pi\alpha/\beta$: this is why the Sommerfeld enhancement is often referred as a $1/v$ enhancement. On the other hand, $S\to 1$ when $\alpha/\beta \to 0$, as one would expect. 

It should however be noted that the $1/v$ behaviour breaks down at very small velocities. The reason is that the condition for neglecting the Yukawa part of the potential is that the kinetic energy of the collision should be much larger than the boson mass $\mV$ times the coupling constant $\alpha$, i.e., $\mdm\beta^2 \gg \alpha\mV$, and this condition will not be fulfilled for very small values of $\beta$. This is also evident if we expand the potential in powers of $x=\mV r$; then, neglecting terms of order $x^2$ or smaller, the Schr\"odinger equation can be written as (the prime denotes the derivative with respect to $x$):
\begin{equation}
\psi''+\frac{\alpha}{\eps}\frac{\psi}{x}=\left(-\frac{\beta^2}{\eps^2}+\frac{\alpha}{\eps}\right)\psi,
\label{eq:sch2}
\end{equation} 
having defined $\eps=\mV/\mdm$.
The Coulomb case is recovered for 
 $\beta^2 \gg \alpha \eps$, or exactly the condition on the kinetic energy stated above. It is useful to define 
$\beta^* \equiv \sqrt{{\alpha \mV}/{\mdm}}$
such that $\beta\gg\beta^*$ is the velocity regime where the Coulomb approximation for the potential is valid.

Another simple, classical interpretation of this result is the following. The range of the Yukawa interaction is given by $R\simeq \mV^{-1}$. Then the crossing time scale is given by $t_{cross}\simeq R/v\simeq 1/\beta \mV$. On the other hand, the dynamical time scale associated to the potential is $t_{dyn}\simeq\sqrt{R^3 \mdm/\alpha}\simeq\sqrt{\mdm/\alpha \mV^3}$. Then the condition $\beta\gg \beta^*$ is equivalent to $t_{cross} \ll t_{dyn}$, i.e., the crossing time should be much smaller than the dynamical time-scale.
Finally, we note that since in the Coulomb case $S \sim 1/\beta $ for $\alpha \gg \beta$, the region where the Sommerfeld enhancement actually has a $1/v$ behaviour is $\beta^*\ll \beta\ll \alpha$. It is interesting to notice that this region does not exist at all when $m \lesssim \mV/\alpha $.

The other interesting regime to examine is $\beta \ll \beta^*$. Following the discussion above, this corresponds to the potential energy dominating over the kinetic term. Referring again to the form (\ref{eq:sch2}) for $x\ll 1 $ of the Schr\"odinger equation, this  becomes:
\begin{equation}
\psi''+\frac{\alpha}{\eps}\frac{\psi}{x}=\frac{\alpha}{\eps}\psi.
\label{eq:sch3}
\end{equation} 
The positiveness of the right-hand side of the equation points to the existence of bound states. In fact, this equation has the same form as the one describing the hydrogen atom. Then bound states exist when $\sqrt{\alpha/\eps}$ is an even integer, i.e. when:
\begin{equation}
\mdm = {4\mV} n^2 / {\alpha}, \qquad n=1,\, 2, \dots
\label{eq:bound}
\end{equation}
From this result,  we expect that the Sommerfeld enhancement will exhibit a series of resonances for specific values of the particle mass spaced in a $1:4:9:...$ fashion. The behaviour of the cross section close to the resonances can be better understood by approximating the electroweak potential by a well potential, for example:
$V(r) = - \alpha \mV \theta(R-r),$
where $R=\mV^{-1}$ is the range of the Yukawa interaction, and the normalization is chosen so that the well potential roughly matches the original Yukawa potential at $r=R$. The external solution satisfying the boundary conditions at infinity is simply an incoming plane wave, $\psi_{out}(r) \propto e^{i k_{out} r}$, with $k_{out} = \mdm\beta$. The internal solution is:
$\psi_{in}(r) = A e^{i k_{in} r} + B e^{-i k_{in} r},$
where $k_{in} = \sqrt{k_{out}^2+\alpha\mdm\mV}\simeq \sqrt{\alpha\mdm\mV}$ (the last approximate equality holds because $\beta\ll \beta^*$). The coefficients $A$ and $B$ are as usual obtained by matching the wave function and its first derivative at $r=R$; then the enhancement is found to be:
\begin{equation}
S=\left[{\cos^2 k_{in}R+\frac{k_{out}^2}{k_{in}^2}\sin^2 k_{in}R}\right]^{-1}.
\end{equation}
When $\cos k_{in} R =0$, i.e., when $\sqrt{\alpha \mdm/\mV} = (2n+1)\pi/2$, the enhancement assumes the value $k_{in}^2/k_{out}^2\simeq {\beta^*}^2/\beta^2\gg 1$. This is however cut off by the finite width of the state.

 In summary, the qualitative features that we expect to observe are\\
i) at large velocities ($\beta \gg \alpha$) there is no enhancement, $S\simeq 1$; \\
ii) in the intermediate range $\beta^*\ll \beta \ll \alpha$, the enhancement goes like $1/v$: $S\simeq \pi\alpha/\beta$, this value being independent of the particle mass; \\
 iii) at small velocities ($\beta\ll \beta^*$), a series of resonances appear, due to the presence of bound states. Close to the resonances, $S\simeq (\beta^*/\beta)^2$. In this regime, the enhancement strongly depends on the particle mass, because it is this that determines whether we are close to a resonance or not. Similar results have been independently obtained in Ref. \cite{MarchRussell:2008tu}.

\begin{figure}
	\begin{center}
	\includegraphics[width=0.45\textwidth,  keepaspectratio]{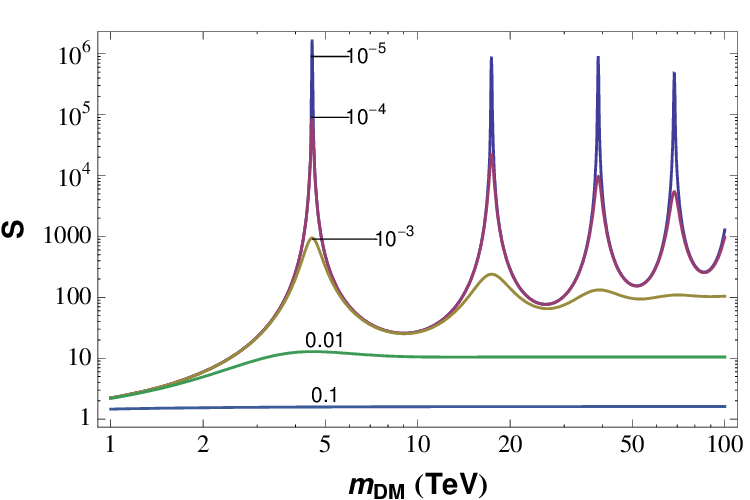}
\caption{Sommerfeld enhancement $S$ as a function of the dark matter particle mass $\mdm$, for different values of the particle velocity. Going from bottom to top $\beta=10^{-1},\,10^{-2},\,10^{-3},\,10^{-4},\,10^{-5}$.}
	\label{fig:som1}
         \end{center}
\end{figure}

We show the result of the numerical integration of Eq. (\ref{eq:sch}) in Figure \ref{fig:som1}, where we plot the enhancement $S$ as a function of the particle mass $\mdm$, for different values of $\beta$. We choose specific values of the boson mass $\mV=90\,\GeV$ and of the gauge coupling $\alpha=\alpha_2\simeq1/30$. These values correspond to a particle interacting through the exchange of a Z boson.

We note however that, as can be seen by the form of the equation, the enhancement depends on the boson mass only through the combination $\eps = \mV/\mdm$, so that a different boson mass would be equivalent to rescaling the abscissa in the plot. Moreover, the evolution of the wave function only depends on the two quantities $\alpha/\eps$ and $\beta/\eps$, so that a change $\alpha\to\alpha'$ in the gauge coupling would be equivalent to:
$\beta \to \beta' = \frac{\alpha'}{\alpha}\beta, \,  \,
\eps  \to \eps'   = \frac{\alpha'}{\alpha}\eps.$
This shows that Fig. \ref{fig:som1} does indeed contain all the relevant information on the behaviour of the enhancement $S$. 

We see that the results of the numerical evaluation agree with our qualitative analysis above.  When $\beta=10^{-1}$ (bottom curve), we are in the $\beta>\alpha\simeq 3\times 10^{-2}$ regime and there is basically no enhancement. The next curve $\beta=10^{-2}$ is representative of the $\beta\gtrsim\beta^*$ regime, at least for $\mdm$ larger than a few TeV.
The enhancement is constant with the particle mass and its value agrees well with the expected value $\pi\alpha/\beta\simeq10$. The drop of the enhancement in the mass region below $\sim 3\ \TeV$ is due to the fact that here $\beta\lesssim\beta^*$, and that there are no resonances for this value of the mass. Decreasing $\beta$ again (top three curves, corresponding to $\beta = 10^{-3},\,10^{-4},\,10^{-5}$ from bottom to top) we observe the appearance of resonance peaks. The first peak occurs for $\mdm=\bar\mdm=4.5\,\TeV$, so that expression (\ref{eq:bound}) based on the analogy with the hydrogen atom overestimates the peak position by  a factor 2. However, the spacing between the peaks is as expected, going like $n^2$, as the next peaks occur roughly at $\mdm=4,\,9,\,16\,\bar\mdm$. The height of the first peak agrees fairly well with its expected value of $(\beta^*/\beta)^2$. The other peaks are damped; this is particularly evident for $\beta=10^{-3}$, and in this case it is due to the fact that  $\beta^*$ decreases as $\mdm$ increases, so that for $\mdm\sim 100\,\TeV$ we return to the non-resonant, $1/\beta$ behaviour, and the enhancement takes the constant value $\pi\alpha/\beta\simeq100$.
\begin{figure}
	\begin{center}
	\includegraphics[width=0.45\textwidth,  keepaspectratio]{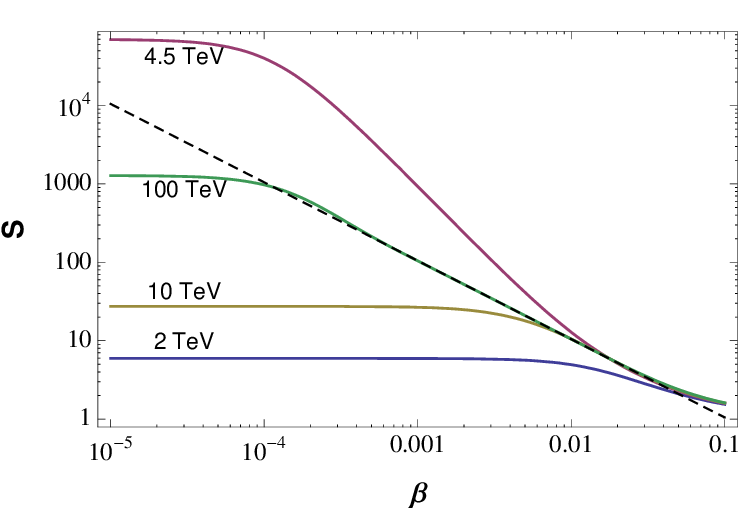}
	\includegraphics[width=0.45\textwidth,  keepaspectratio]{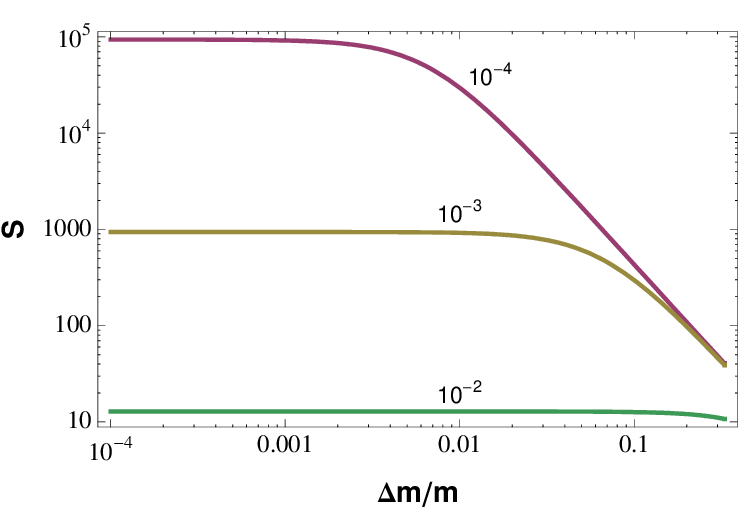}
          \caption{Top panel: Sommerfeld enhancement $S$ as a function of the particle velocity $\beta$ for different values of the dark matter mass. From bottom to top: $\mdm= 2,\,10,\,100,\,4.5\,\TeV$,  the last value corresponding to the first resonance in Fig. \ref{fig:som1}. The black dashed line shows the $1/v$ behaviour that is expected in the intermediate velocity range (see text for discussion).  Bottom panel: Sommerefeld enhancement $S$ as a function of the relative distance from the first resonance shown in Fig. \ref{fig:som1}, occurring at $m\simeq 4.5\,\TeV$, for different values of $\beta$. From top to bottom: $\beta=10^{-4},\,10^{-3},\,10^{-2}$.}
	\label{fig:som2}
         \end{center}
\end{figure}

Complementary information can be extracted from the analysis of the upper panel of Fig. \ref{fig:som2}, where we plot the Sommerfeld enhancement as a function of $\beta$, for different values of the particle mass. Far from the resonances, the enhancement factor initially grows as $1/\beta$ and then saturates to some constant value. This constant value can be estimated by solving the Schr\"odinger equation with $\beta=0$. We find that a reasonable order of magnitude estimate is given by $S_{max}\sim 6\alpha/\eps$;  the corresponding value of $\beta\sim0.5\eps$.   The $1/\beta$ behaviour holds down to smaller velocities for larger particle masses, leading to larger enhancement factors. However, when the particle mass is close to a resonance, $S$ initially grows like $1/\beta$ but at some point the $1/\beta^2$ behaviour "turns on", leading to very large values of the boost factor, until this also saturates to some constant value.

It is clear from the discussion until this point that the best hope for obtaining a large enhancement comes from the possibility of the dark matter mass lying close to a resonance; for the choice of parameter used above this would mean $\mdm\simeq \bar \mdm \simeq 4.5\,\TeV$. However, one could be interested in knowing how close the mass should be  to the center of the resonance in order to obtain a sizeable boost in the cross-section. In order to understand this, we show in Fig. \ref{fig:som2} the enhancement as a function of $\mu\equiv |\mdm-\bar\mdm|/\mdm$, i.e., of the fractional shift from the center of the resonance. Clearly, for $\beta\le10^{-3}$, a boost factor of $\gtrsim 100$ can be obtained for $\mu\lesssim 0.2$, i.e., for deviations of up to 20\% from $\bar \mdm$, corresponding to the range between 3.5 and 5.5 TeV. This is further reduced to the 4 to 5 TeV range if one requires $S\gtrsim 10^3$.

\section{The leptonic branching ratio}
\label{sec:lept}
The relevance of the Sommerfeld enhancement for the annihilation of supersymmetric particles was first  pointed out  in Refs ~\cite{hisano,Hisano05}, in the context of the minimal supersymmetric standard model where the neutralino is the lightest supersymmetric particle. A wino-like or higgsino-like neutralino would interact with the W and Z gauge bosons due to its SU(2)$_L$ nonsinglet nature. In particular, the wino $\tilde W^0$ is the neutral component of a SU(2)$_L$ triplet , while the higgsinos $(\tilde H_1^0,\, \tilde H_2^0)$ are the neutral components of two SU(2)$_L$ doublets. The mass (quasi-) degeneracy between the neutralino and the other components of the multiplet leads to transitions between them, mediated by the exchange of weak gauge bosons; this gives rise to a Sommerfeld enhancement at small velocities. On the other hand, the bino-like neutralino being a SU(2) singlet, would not experience any Sommerfeld enhancement, unless a mass degeneracy with some other particle is introduced into the model.

The formalism needed to compute the enhancement when mixing among states is present is slightly more complicated than the one described above, but the general strategy is the same. As shown in the paper by Hisano et al. \cite{Hisano05} through direct numerical integration of the Schr\"odinger equation, the qualitative results of the previous section still hold: for dark matter masses $\gtrsim 1\,\TeV$, a series of resonances appear, and the annihilation cross section can be boosted by several order of magnitude.

\begin{figure}
	\begin{center}
\includegraphics[width=0.45\textwidth,  keepaspectratio]{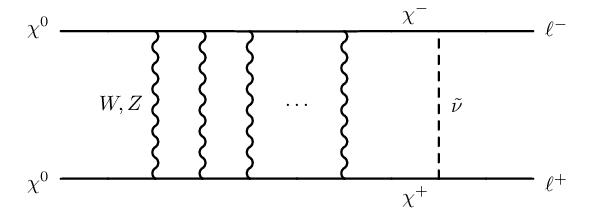}
\caption{Diagram describing the annihilation of two neutralinos into a charged lepton pair, circumventing helicity suppression.}
\label{fig:feyn}
         \end{center}
\end{figure}

An interesting feature of this ``multi-state'' Sommerfeld effect is the possibility of boosting the cross section for some annihilation channels more than others. This happens when one particular annihilation channel is very suppressed (or even forbidden) for a given two-particle initial state, but not for other initial states. This can be seen as follows. The general form for the total annihilation cross section after the enhancement has been taken into account is
\begin{equation}
\sigma v =N \sum_{ij} \Gamma_{ij} d_i(v) d_j^*(v),
\end{equation}
where $N$ is a multiplicity factor, $\Gamma_{ij}$ is the absorptive part of the action, responsible for the annihilation, the $d_i$ are coefficients describing the Sommerfeld enhancement, and the indices $i,j$  run over the possible initial two-particle states. Let us consider for definiteness the case of the wino-like neutralino: the possible initial states are $\{\chi^0\chi^0,\,\chi^+\chi^-\}$. The neutralino and the chargino are assumed to be quasi-degenerate, since they are all members of the same triplet. What we will say can anyway be easily generalized to the case of the higgsino-like neutralino. 
Let us also focus on two particular annihilation channels: the $W^+W^-$ channel and the $e^+e^-$ channel. It can be assumed that, close to a resonance, $d_1\sim d_2$. This can be inferred for example using the square well approximation as  in Ref. \cite{Hisano05}, where it is found that, in the limit of small velocity, $d_{1}\simeq\sqrt2(\cos\sqrt2 p_c)^{-1}-\sqrt2(\cosh p_c)^{-1}$ 
and $d_{2}\simeq(\cos\sqrt2 p_c)^{-1}+2(\cosh p_c)^{-1}$, where $p_c\equiv\sqrt{2\alpha_2 \mdm/m_W}$. The elements of the $\Gamma$ matrix for the annihilation into a pair of $W$ bosons are $\sim \alpha^2_2/m_\chi^2$, so that we can write the following order of magnitude estimate:
\begin{equation}
\sigma v (\chi^0\chi^0\to W^+W^-) \sim |d_1|^2 \frac{\alpha^2_2}{m_\chi^2}.
\end{equation}
On the other hand, the non-enhanced neutralino annihilation cross section to an electron-positron pair $\Gamma_{22}\sim \alpha_2^2 m_e^2/m_\chi^4$, so that it is suppressed by a factor $(m_e/m_\chi)^2$ with respect to the gauge boson channel. This is a well-known general feature of neutralino annihilations to fermion pairs and is due to the Majorana nature of the neutralino. The result is that  all low velocity neutralino annihilation diagrams to fermion pairs have amplitudes proportional to the final state fermion mass.  The chargino annihilation cross section to fermions, however, does not suffer from such an helicity suppression, so that it is again $\Gamma_{11}\sim \alpha_2^2/m_\chi^2\gg \Gamma_{22}$. Then:
\begin{equation}
\sigma v (\chi^0\chi^0\to e^+e^-) \sim |d_1|^2 \frac{\alpha^2_2}{m_\chi^2}.
\end{equation}
Then we have that, after the Sommerfeld correction, the neutralino annihilates to W bosons and to $e^+e^-$ pairs (and indeed to all fermion pairs) with similar rates, apart from $O(1)$ factors. This means that while the $W$ channel is enhanced by a factor $|d_1|^2$, the electron channel is enhanced by a factor $|d_1|^2 m_\chi^2/m_e^2$. The reason is that the annihilation can proceed through a ladder diagram like the one shown in Fig.~\ref{fig:feyn}, in which basically the electron-positron pair is produced by annihilation of a chargino pair close to an on-shell state. This mechanism can be similarly extended to annihilations to other charged leptons, neutrinos or quarks.

\section{CDM substructure: enhancing the Sommerfeld boost}

There is a vast reservoir of clumps in the outer halo where they spend most of their time. Clumps should survive perigalacticon passage  over  a fraction   (say $\nu$)  of an orbital time-scale,  $ t_d =r/v_{r},$ where $v_{r}$ is the orbital velocity (given by   $v_r^2=GM/r)$. It is reasonable to assume that the survival probability is a function of the ratio between $t_d$ and the age of the halo $t_H$, and that it vanishes for $t_d \to 0$. Thus, at linear order in the (small) ratio $t_d/t_H$, a first guess at the clump mass fraction as a function of galactic radius would be $f_{clump} \propto t_d. $  We conservatively adopt the clump mass fraction $\mu_{cl}=\nu rv_{r}^{-1} t_H^{-1} $ with $\nu =0.1-1$. This gives a crude but adequate fit  to  the highest resolution simulations, which find that the outermost halo has a high clump survival fraction, but that near the sun only 0.1-1 \%  survive \cite{springel}.  In the innermost galaxy, essentially all clumps are destroyed. 
  
Suppose the clump survival fraction  $S(r) \propto f_{clump} \propto r^{3/2} $ to zeroth order. The annihilation flux  is proportional  to $\rho^2 \times {\rm Volume} \times  S(r) \propto S(r)/r$.  This suggests we  should expect to find an appreciable   gamma ray flux from  the outer galactic halo. It should be quasi-isotropic with a $\sim$10\% 
offset from the centre of the distribution. The flux from the Galactic Centre would be superimposed on this. High resolution simulations demonstrate that clumps account for as much luminosity as the uniform halo \cite{stadel}, \cite{springela}.
However much of  the soft lepton excess from the inner halo will be suppressed  due to the clumpiness being much less in the  inner galaxy.

We see from  the numerical simulations of our halo, performed at a mass resolution of $1000\rm M_\odot$ 
that the subhalo contribution to the annihilation luminosity scales as $M_{min} ^ {-0.226}$ \cite{springela}. For $M_{min}=10^5 M_\odot$, this roughly equates the contribution of the smooth halo at $r=200$ kpc from the center. This should continue  down to the minimum subhalo mass. We take the latter to be $10^{-6} \rm M_\odot $ clumps, corresponding the damping scale of a bino-like neutralino \cite{Hofmann:2001bi,Loeb:2005pm}. We consider this as representative of the damping scale of neutralino dark matter, although it should be noted that the values of this cutoff for a general weakly interacting massive particle (WIMP) candidate can span several orders of magnitude, depending on the details of the underlying particle physics model \cite{Profumo:2006bv,Bringmann:2009vf}. It should also be taken into account that the substructure is a strong function of galactic radius. Since the dark matter density drops precipitously outside the solar circle 
(as $r^{-2}$), the clump contribution to boost  is important in the solar neighbourhood. However absent any Sommerfeld boost,  it amounts only to a factor of order unity. Incidentally the simulations show that most of the luminosity occurs in the outer parts of the halo  \cite{springela} and that the boost here due to substructure is large,  typically  a factor of  230 at $r_{200}.$ 
 
However there is another effect of clumpiness, namely low internal velocity dispersion. In fact, the preceding discussion greatly underestimates the clump contribution to the annihilation signal. This is because the coldest substructure survives clump destruction albeit  on microscopic scales. Within the clumps, the velocity dispersion $\sigma$  initially is low.    
Thus, the annihilation cross section is further enhanced by the Sommerfeld effect in the coldest surviving substructure. We now estimate that including this effect results in a Sommerfeld-enhanced clumpiness boost factor at the solar neighborhood  of $10^4$ to $10^5.$ 

To infer $\sigma$ from the mass $M$  of the clump is straightforward. The scalings can be obtained by combining dynamically self-consistent solutions for the radial dependence of the phase space density in simulated CDM halos  
\cite{Dehnen  and McLaughlin}
 as well as directly from the simulations \cite{vass}
$\rho/\sigma^{\epsilon}\propto r^{-\alpha},$  combined with our ansatz about clump survival that relates minimum clump mass to radius and the argument that marginally surviving clumps have density contrast of order unity. With $\epsilon=3$ and $\alpha = 1.875$ \cite{navarro},
we  infer (for the isotropic case)  that $\sigma \propto \rho^{1/{\epsilon}}r^{\alpha/\epsilon} \simpropto M^{1/4}$.  This is a compromise between the two exact solutions for nonlinear clumps formed from hierarchical clustering of CDM: spherical  ($M\propto r^3$) or Zeldovich pancakes ($M\propto r$), and is just the self-similar scaling  limiting value.
The numerical simulations of \cite{springel} suggest a scaling $M_{sub}\propto v_{max}^{3.5}$ down to the resolution limit of $\sim 10^3 \rm M_\odot ,$ somewhat steeper than self-similar scaling. 

So one  can combine this result with the
previous  scaling to compute the total boost, i.e., taking into account both the clumpiness and the Sommerfeld enhancement. We know from the analysis of Springel et al. \cite{springela} that
for a minimum halo mass of $10^{-6}\,M_\odot$ the luminosity of the subhalo component should more or less equate to that of the smooth halo at the galactocentric radius, i.e. $L^0_{sh}\simeq L^0_{sm}$ at $r=8\,\kpc$, where the superscript 0 stands for the luminosity in the absence of any Sommerfeld correction. Thus the boost factor  with respect to a smooth halo is of order unity, after the presence of subhalos is taken in consideration. Next we take into account the Sommerfeld enhancement. The velocity dispersion  in the halo is $\beta \sim 10^{-3}$, while the velocity dispersion in the subhalos is $\beta \sim 10^{-5}$ for a $10^5\,M_\odot$ clump, and can be scaled down to smaller clumps using the $\sigma\simpropto M^{1/4}$ relation. From the discussion in sec. \ref{sec:som} and in particular from Figs. \ref{fig:som1} and \ref{fig:som2} it appears that, if the dark matter mass is $\lesssim 10\,\TeV$ and far from the resonance occurring for $m\simeq 4.5\,\TeV$: (1) the Sommerfeld enhancement is the same for the halo and for the subhalos, since it has already reached the saturation regime; (2) it is of order 30 at most, so that the resulting boost factor still falls short by at least one order of magnitude with respect to the value needed to explain the PAMELA data. On the other hand, if the dark matter mass is close to its resonance value, then a larger value of the boost can be achieved inside the cold clumps, since (1) the enhancement is growing like $1/v^2$ and (2) it is saturating at a small value of $\beta$. Referring for definiteness to the top curve in the top panel of Fig. \ref{fig:som2} ($m=4.5\,\TeV$), one finds $S\simeq 10^4-10^5$ for all clumps with mass $M\lesssim 10^9\,M_\odot$ (that is roughly the mass of the largest clumps) while the smooth halo is enhanced by a factor 1000. 
Then the net result is that the boost factor is of order $10^4-10^5$ and is mainly due to the Sommerfeld enhancement in the cold clumps (the enhancement in the diffuse halo only contributing a fraction  1-10\%). Of course the details will be model dependent; it should also be stressed that the enhancement strongly depends on the value of the mass when this is close to the resonance.

\section{Discussion}

In the previous section we have shown how it is possible to get a boost factor of order $10^4-10^5$ for a dark matter particle mass of order $4.5\,\TeV$. This is tantalizing because this is roughly the value one needs to explain the PAMELA data for a dark matter candidate with this given mass, as can be inferred by analysis of Fig. 9 of Ref \cite{Cirelli:2008pk}. Although we have made several approximations concerning the clump distribution and velocity, it should be noted that our results still hold as long as the majority of the clumps are very cold ($\beta\lesssim 10^{-4}$) because this is the regime in which the enhancement becomes constant. The saturation of the Sommerfeld effect also plays a crucial role in showing that the very coldest  clumps are unable to contribute significantly  to the required boost factor if the dark matter mass is not close to one of the Sommerfeld resonances. Because of saturation below $\beta\sim 10^{-4}$, the Sommerfeld boost is insensitive to extrapolations beyond the currently resolved  scales in simulations. Note however that the precise value for the dark matter particle mass is uncertain because of such model-dependent assumptions as the adopted  
mass-splitting, the multiplet nature of the supersymmetric particles, and the possibility of different couplings, weaker than weak.

The model presented here does not pose any problem from the point of view of the high energy gamma-ray emission from the centre of the galaxy, since very few clumps are presents in the inner core and thus there is no Sommerfeld enhancement. Thus there is no possibility of violating the EGRET or HESS observations of the galactic center or ridge, contrary to what is argued in Ref. \cite{Bertone08}. There is a potential problem however with gamma ray production beyond the solar radius out to the outer halo. From \cite{springela}, the simulations are seen to yield an additional enhancement due to clumpiness alone  above  $10^5\rm M_\odot$ of around 80\% at $r_{200}$ in the annihilation luminosity. Extrapolating to earth mass clumps, the enhancement is 230 in the annihilation luminosity
at the same radius.  This is what a distant observer would see. 
The incorporation of the Sommerfeld factor would greatly amplify this signal by $S\sim 10^4-10^5.$ 

\begin{figure}
	\begin{center}
	\includegraphics[width=0.45\textwidth,  keepaspectratio]{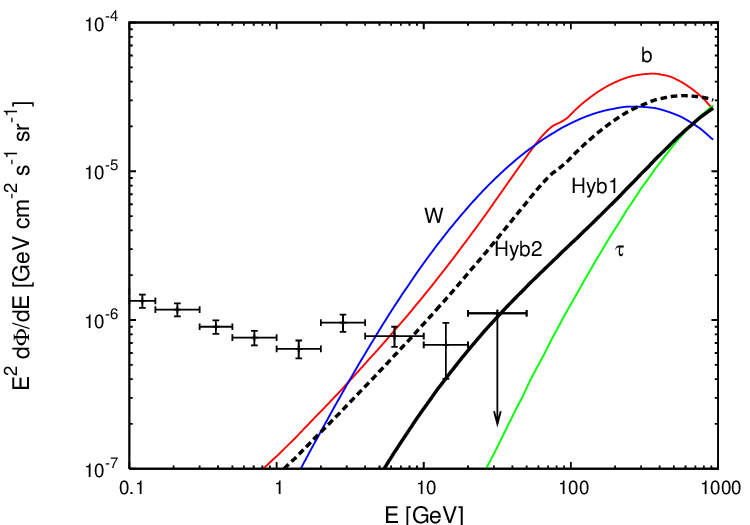}
          \caption{Contribution to the diffuse galactic photon background from the annihilation of a 5 TeV dark matter particle, for different channels, when both clumpiness and the Sommerfeld enhancement in cold clumps are taken into account, compared with the measurements of the diffuse gamma background from EGRET \cite{Strong:2004ry}.  The label ``Hyb1'' (solid black line) stands for a hybrid model in which the dark matter annihilates to $\tau$ leptons 90\% of the time and to $W$ pairs the rest of the time. The label ``Hyb2'' (dashed black line) stands for a model in which the dark matter annihilates to leptons and quarks only, with the same cross-section apart from color factors. The latter could be realized through the circumvention of helicity suppression.}
	\label{fig:diff}
         \end{center}
\end{figure}

The expected flux that would be observed by looking in a direction far from the galactic center can be readily estimated. Assuming an effective cross section $\sigma v= 3\times 10^{-22}$ cm$^3$ s$^{-1}$, corresponding to a Sommerfeld boost of $10^4$ on top of the canonical value of the cross section times velocity, the number of annihilations on the line of sight is roughly  $4\times 10^{-9} (\mdm/\TeV)^{-2}$ cm$^{-2}$ s$^{-1}$. We have assumed a Navarro-Frenk-White (NFW) profile. The effect of the clumpiness is still not included in this estimate. Following the results of the simulation in Ref. \cite{springela}, this value should be multiplied by a factor $\sim 200.$ 
Convolving with the single annihilation spectrum of a 5 TeV dark matter particle yields  the flux shown in Fig. \ref{fig:diff}. There we show the spectrum that would be produced if the dark matter particle would annihilate exclusively either to $W$ bosons, b quarks or $\tau$ leptons (blue, red and green curves, respectively). We also consider a candidate that annihilates to $\tau$ leptons 90\% of the time and to $W$s the remaining 10\% of the time (model ``Hyb1'') and a candidate that annihilates only to quarks and leptons, with the same cross section apart from color factors (model ``Hyb2'').

The gamma ray signal  mostly originates from the outer halo and should be detectable as an almost isotropic hard gamma-ray background. Candidates annihilating to heavy quarks or to gauge bosons seem to be excluded by EGRET. On the other hand,  a dark matter particle annihilating to $\tau$ leptons is compatible with the measurements of EGRET at these energies \cite{Strong:2004ry}, and within the reach of FERMI.

There are however at least two reasons that induce  significant uncertainty into any estimates. 
 Firstly,  the halo density profile in the outer galaxy may be  substantially steeper than is inferred from an NFW profile, as current models  
are  best fit by an Einasto profile \cite{gao},
$\rho(r)\propto \exp[(-2/\alpha ((r/r_s)^\alpha - 1)],$ as opposed to the asymptotic NFW profile $\rho(r)\propto r^{-3}$. Using the Einasto profile  yields at least a  10\% reduction.  
Another possibility is to use a Burkert profile \cite{Burkert:1995yz}, that gives a better phenomenological description of the dark matter distribution inside the halo, as it is inferred by the rotation curves of galaxies \cite{Gentile:2004tb,Salucci:2007tm}. Using a Burkert profile, the flux is reduced by a factor 3.
Secondly, and  more importantly,  the subhalos are much less concentrated at greater distances from the Galactic Centre \cite{diemand}. 
These effects should substantially reduce the gamma ray contribution from the outer halo.  A future application will be to evaluate the extragalactic diffuse gamma ray background where the evolution of clumpiness with redshift should play an interesting role in producing a possible spectral feature in the isotropic component. Note that the annihilation rate originating from very high redshift subhalo substructure and clumpiness near the neutralino free-streaming scale \cite{Kamionkowski:2008gj} is mostly suppressed due to the saturation of the Sommerfeld effect that we described above. 

Because of the  saturation of the Sommerfeld boost, it should be possible to focus future simulations on improved modelling of the radial profiles and concentrations of substructures in the outer halo. It is these that contribute significantly to the expected diffuse gamma if our interpretation of the PAMELA
and  the ATIC data,  and in particular the required normalisation and hence boost, is correct. Of course, there are other possible explanations of the  high energy positron data, most notably the flux from a local pulsar \cite{ahar, yuks,Hooper:2008kg} that has recently been detected as a TeV gamma ray source.

An interesting consequence of the model proposed here is the production of synchrotron radiation emitted by the electrons and positrons produced in the dark matter annihilations, similar to the one that is possibly the cause of the observed ``WMAP haze'' \cite{Hooper:2007kb,Cumberbatch:2009ji}. For a TeV candidate, this synchrotron emission would be visible in the $\nu \gtrsim 100$~GHz frequency region. This region will be probed by the Planck mission; the synchrotron radiation would then give rise to a  galactic foreground ``Planck haze'' in the microwave/far infrared part of the spectrum. This quasi-isotropic high frequency synchrotron component will be an additional source of B-mode
foregrounds that will need to be incorporated into proposed attempts to disentangle any primordial B-mode component in the cosmic microwave background. Another interesting application would be to look at the gamma-ray emission from specific objects, like the Andromeda Galaxy (M31). M31 has been observed in the relevant energy range by the CELESTE and HEGRA atmospheric Cherenkov telescopes, and limits on the partial cross section to photons, in the absence of boost, were obtained in Ref.   \cite{Mack:2008wu}.

Finally, we note that  in Sec. \ref{sec:lept} we have described a mechanism that can enhance the production of leptons (especially light leptons) in neutralino dark matter annihilations, making the leptonic channel as important as the gauge boson channel. A dark matter candidate annihilating mainly into leptons can simultaneously fit the PAMELA positron and antiproton data, owing to the fact that no antiproton excess is produced. The enhancement of the lepton branching ratio can possibly alleviate the problem of antiproton production following neutralino annihilation into a pair of gauge bosons. It should however be noted that the mechanism in question also enhances the quark channel in a similar way, thus introducing an additional source of antiprotons. It would thus be desirable to suppress in some way the quark annihilation channel. This could be realised in a variation of the above mentioned mechanism, if the lightest neutralino is quasi-degenerate in mass with the lightest slepton $\tilde l$; this is what happens for example in the $\tilde \tau$ coannihilation region. In this case, the Sommerfeld enhancement would proceed through the creation of an intermediate $\tilde l^+\tilde l^-$ bound state that would subsequently annihilate to the corresponding standard model lepton pair, without producing any (tree-level) quark.
This points to the necessity of further investigating different models in order to assess if the boost in the leptonic branching ratio is indeed compatible with the PAMELA data.

\section*{Acknowledgements} 
The authors would like to thank J. Beacom, M. Cirelli, D. Cumberbatch, J. March-Russell, A. Masiero, A. Strumia and S. West for useful discussions. ML is supported by the Istituto Nazionale di Fisica Nucleare.

\end{document}